\documentclass[fleqn,usenatbib]{mnras}

\usepackage{newtxtext,newtxmath}

\usepackage[T1]{fontenc}
\usepackage{ae,aecompl}
\usepackage[para,online,flushleft]{threeparttable}
\newcommand{\Msun} {M$_{\odot}$}

\newcommand{\Porb}{P$_{\rm orb}$}

\usepackage{graphicx}	
\usepackage{amsmath}	
\usepackage{amssymb}	

\title[LISA verification binaries]{LISA verification binaries with updated distances from {\sl Gaia} Data Release 2}

\author[T. Kupfer et al.]{
T. Kupfer$^{1}$\thanks{E-mail: tkupfer@caltech.edu},
V. Korol$^2$, 
S. Shah$^3$,
G. Nelemans$^{4,5}$,
T. R. Marsh$^6$,
G. Ramsay$^7$,
\newauthor
P. J. Groot$^4$,
D. T. H Steeghs$^6$,
E. M. Rossi$^2$,
\\
$^1$Division of Physics, Mathematics and Astronomy, California Institute of Technology, Pasadena, CA 91125, USA\\
$^2$Leiden Observatory, Leiden University, PO Box 9513, 2300 RA, Leiden, the Netherlands\\
$^3$Albert-Einstein-Institut, Max-Planck-Institut f\'ur Gravitationsphysik, D-30167 Hannover, Germany\\
$^4$Department of Astrophysics/IMAPP, Radboud University Nijmegen, P.O.Box 9010, 6500 GL, Nijmegen, The Netherlands \\
$^5$Institute of Astronomy, KU Leuven, Celestijnenlaan 200D, B-3001 Leuven, Belgium \\
$^6$Department of Physics, University of Warwick, Coventry CV4 7AL, UK\\
$^7$Armagh Observatory and Planetarium, College Hill, Armagh, BT61 9DG,UK
}

\date{Accepted XXX. Received YYY; in original form ZZZ}

\pubyear{2015}

\begin{document}
\label{firstpage}
\pagerange{\pageref{firstpage}--\pageref{lastpage}}
\maketitle

\begin{abstract}
Ultracompact binaries with orbital periods less than a few hours will dominate the gravitational wave signal in the mHz regime. Until recently, 10 systems were expected have a predicted gravitational wave signal strong enough to be detectable by the Laser Interferometer Space Antenna (\emph{LISA}), the so-called `verification binaries'. System parameters, including distances, are needed to provide an accurate prediction of the expected gravitational wave strength to be measured by \emph{LISA}. Using parallaxes from {\sl Gaia} Data Release 2 we calculate signal-to-noise ratios (SNR) for $\approx$50 verification binary candidates. We find that 11 binaries reach a SNR$\geq$20, two further binaries reaching a SNR$\geq$5 and three more systems are expected to have a SNR$\approx$5 after four years integration with \emph{LISA}. For these 16 systems we present predictions of the gravitational wave amplitude ($\mathcal{A}$) and parameter uncertainties from Fisher information matrix on the amplitude ($\mathcal{A}$) and inclination ($\iota$).


\end{abstract}





\section{Introduction}\label{sec:intro}
%
%

The Laser Interferometer Space Antenna, (\emph{LISA}) will be the first gravitational wave observatory in space (\citealt{ama17}). Operating in the low frequency part of the gravitational wave (GW) spectrum ($10^{-4} - 1$ Hz), \emph{LISA} has been selected as ESA's third large mission of the Cosmic Vision Program\footnote{launch planned between 2030-2034}. Amongst many other astrophysical sources, \emph{LISA} will allow us to observe millions of ultracompact (Galactic) binaries (UCBs) with orbital periods (\Porb) shorter than a few hours (\citealt{ama17}) from which we will be able to individually resolve several thousands (e.g. \citealt{nel04}; \citealt{rui09}; \citealt{rui10, yu10, liu10, sha12, nis12, nel13, lit13, kre17, bre18}). As indicated by their tight orbits, these systems are composed of degenerate stellar remnants, such as white dwarfs, neutron stars or stellar-mass black holes. Up to now several such sources have been detected in the electromagnetic (EM) bands. These include detached \citep{bro16a} and semi-detached double white dwarfs  (the latter called AM\,CVn type binaries; \citealt{sol10}), and semi-detached white dwarf-neutron star binaries (so-called ultracompact X-ray binaries; \citealt{nel10a}) and double neutron stars \citep{lyn04}. 

A subset of the known UCBs have orbital periods that lie in the \emph{LISA} band and these will be individually detected due to their strong GW signals. These \emph{LISA}-guaranteed sources are termed `verification binaries' with some being expected to be detected on a timescale of weeks or a few months \citep{str06}.  Therefore, they are crucial in facilitating the functional tests of the instrument and maximising the scientific output of \emph{LISA}. So far we know of ten such systems, most of them being semi-detached AM\,CVn type: HM Cnc, V407 Vul, ES Cet, AM CVn, SDSS J190817.07+394036.4 (SDSS\,J1908), HP Lib, CR Boo, and V803 Cen \citep{str04,ram05,esp05,roe06,roe07c,kup15,gre18}.
The remaining two are detached binary white dwarf systems: SDSS\,J065133.34+284423.4 (SDSS\,J0651) and SDSS\,J093506.92+441107.0 (SDSS\,J0935) \citep{bro11,kil14}. 

Predicting the gravitational wave strain depends on the masses of the binary components, which, together define the chirp mass (defined in \S \ref{sec:snrcal}), the orbital inclinations of the systems and their distance. Masses can be obtained, within limits, from optical spectroscopy and photometry, combined with the Roche-lobe geometry. In favourable situations, such as eclipsing systems, the orbital inclination can be determined from time-resolved spectroscopy and photometry (e.g. \citealt{bro11}), but it is generally rather poorly constrained. So far, distances remained the largest uncertainty. Only five of the known 52 semi-detached AM\,CVn type systems have HST-based parallaxes \citep{roe07c}: AM CVn, HP Lib, CR Boo, V803 Cen and GP Com. Ground-based parallaxes were derived for AM CVn (C.Dahn, as quoted by \citealt{nel04}), GP Com \citep{tho03} and V396 Hya \citep{tho08}. The remaining systems have distance estimates based on the comparison of model fluxes with the observations. These are considered highly uncertain as they require good knowledge of system parameters such as mass ratios, donor properties and accretion rates. 
Of the detached verification binary candidates, only one (WD\,1242-105; \citealt{deb15}) has a parallax measurement. The remaining systems have indirect distance estimates based on the comparison of measured temperatures and surface gravities with stellar models (e.g. \citealt{alt13,ist14,bro16}). 



\begin{table*}
\begin{center}
\caption{Physical properties of the known verification binaries. Masses and inclination angles in brackets are assumed and based on evolutionary stage and mass ratio estimations}
\begin{tabular}{lrrrllll}
\hline
Source                &  $l_{\rm Gal}$   &  $b_{\rm Gal}$   & Orbital period  &  $m_{\rm 1}$ & $m_{\rm 2}$ & $\iota$ & Refs.  \\
                      & (deg)  & (deg)    &  (sec)         & (\Msun) &    (\Msun)  &  ($\deg$)    \\
\hline
 \multicolumn{3}{l}{{\bf AM\,CVn type}}      &     \\
HM Cnc                     &  206.9246  &   23.3952  &   321.529  &   0.55     &  0.27  & $\approx$38 & 1,2\\
V407 Vul                   &   57.7281  &    6.4006  &   569.395  &   [0.8$\pm$0.1]   & [0.177$\pm$0.071]   & [60]  & 3 \\
ES Cet                     &  168.9684  & --65.8632  &   620.21   & [0.8$\pm$0.1]  &  [0.161$\pm$0.064] &  [60] & 4 \\ 
SDSS J135154.46--064309.0  &  328.5021  &   53.1240  &   943.84   &    [0.8$\pm$0.1]     &  [0.100$\pm$0.040]  & [60]  & 5 \\
AM CVn                     &  140.2343  &   78.9382  &  1028.73   &  0.68$\pm$0.06 &  0.125$\pm$0.012  & 43$\pm$2 & 6,7 \\
SDSS J190817.07+394036.4   &   70.6664  &   13.9349  &  1085.7    & [0.8$\pm$0.1]  & [0.085$\pm$0.034]  & 10 - 20 & 8,9  \\
HP Lib                     &  352.0561  &   32.5467  &  1102.70   &  0.49-0.80 & 0.048-0.088 & 26-34 & 10,11 \\
PTF1 J191905.19+481506.2   &   79.5945  &   15.5977  &  1347.35   & [0.8$\pm$0.1] & [0.066$\pm$0.026] & [60] & 12 \\
CXOGBS J175107.6--294037   &  359.9849  &  --1.4108  &  1375.0    & [0.8$\pm$0.1] &  [0.064$\pm$0.026] & [60]  & 13 \\
CR Boo                     &  340.9671  &   66.4884  &  1471.3   & 0.67-1.10 &  0.044-0.088 & 30  & 11,14 \\
V803 Cen                   &  309.3671  &   20.7262  &  1596.4   &  0.78-1.17 & 0.059-0.109 & 12 - 15 & 11,15 \\
     \noalign{\smallskip}
 \multicolumn{3}{l}{{\bf Detached white dwarfs}}         & \\
SDSS J065133.34+284423.4   &  186.9277  &   12.6886  &  765.5   &  0.247$\pm$0.015 &  0.49$\pm$0.02 & ${86.9^{+1.6}_{-1.0}}$  & 16,17 \\
SDSS J093506.92+441107.0   &  176.0796  &   47.3776  & 1188.0   &  0.312$\pm$0.019 &  0.75$\pm$0.24 &  [60]  & 18,19 \\
SDSS J163030.58+423305.7   &   67.0760  &   43.3604  & 2389.8   &  0.298$\pm$0.019 &  0.76$\pm$0.24 & [60]  & 18,20 \\
SDSS J092345.59+302805.0   &  195.8199  &   44.7754  & 3883.7   &  0.275$\pm$0.015 &  0.76$\pm$0.23 & [60] & 18,21 \\
      \noalign{\smallskip}
  \multicolumn{3}{l}{{\bf Hot subdwarf binaries}}      & \\
CD--30$^\circ$11223       &   322.4875  &   28.9379   & 4231.8   &   0.54$\pm$0.02     &  0.79$\pm$0.01  & 82.9$\pm$0.4 & 22 \\
\hline
\end{tabular}
\begin{flushleft}
[1]\citet{str05}, [2]\citet{roe10}, [3]\citet{ram02}, [4]\citet{esp05}, 
[5]\citet{gre18}, [6]\citet{ski99}, [7]\citet{roe06}, [8]\citet{fon11}, [9]\citet{kup15}, [10]\citet{pat02} , [11]\citet{roe07c}, [12]\citet{lev14},[13]\citet{wev16}, [14]\citet{pro97}, [15]\citet{roe07}, [16]\citet{bro11}, [17]\citet{her12}, [18]\citet{bro16}, [19]\citet{kil14}, [20]\citet{kil11a}, [21]\citep{bro10}, [22]\citet{gei13}
\label{tab:system}
\end{flushleft}
\end{center}
\end{table*}

In April 2018, the {\sl Gaia} collaboration released sky positions, parallaxes, and proper motions for more than 1.3 billion sources, with a limiting magnitude of $G = 21$\,mag \citep{gai16,gai18}. Here we present new results on the predicted gravitational wave signal detectable by \emph{LISA} for known Galactic binaries using distances from {\sl Gaia} Data Release 2 (DR2) and the current \emph{LISA} baseline configuration. We calculate updated signal-to-noise (SNR) ratios. For the loudest known verification binaries with SNR$\gtrsim5$, we extract their GW parameter uncertainties using Fisher information matrix. For our analysis we define systems as verification binaries if the binary 1) is detected in the electromagnetic bands and 2) its SNR is $\geq5$ after 4 years of integration which is the nominal mission time for \emph{LISA}. 



\section{The sample of verification binaries}

Observationally, the known sample of 10 verification binaries is strongly biased and incomplete. This sample includes AM CVn, CR Boo, V803 Cen and ES Cet, which were all found as `outliers' in surveys for blue, high-Galactic latitude stars. The selection effects are difficult to accurately quantify and therefore not easy to model. HM\,Cnc and V407\,Vul are the most compact known systems and were discovered during the course of the {\sl ROSAT} All-Sky Survey showing an on/off X-ray profile modulated on a period of 321 and 569 sec respectively \citep{isr99,mot96}. Their ultracompact nature was later confirmed with optical observations \citep{ram02,isr02,ram00,roe10}. SDSSJ1908 was found as a short-period variable in the original Kepler field, where it was targeted as a potential subdwarf B-star pulsator (\citealt{fon11,kup15}). 

Extremely low mass (ELM) white dwarf binaries such as SDSS\,J0651 and SDSS\,J0935 were discovered as part of a colour selected sample of B-type hypervelocity candidates from the Sloan Digital sky survey (SDSS; \citealt{bro16a} and references therein). ELM white dwarfs can be separated efficiently from the bulk of white dwarfs with a colour selection. 

Studies of UCBs have been conducted almost exclusively at high Galactic latitudes and in the Northern hemisphere. It is therefore likely that more `AM CVn'-like systems are awaiting discovery in the Southern hemisphere and at low Galactic latitudes.  
Binary population studies predict that \emph{LISA} will detect several thousand detached and semi-detached double white dwarfs as well as a few tens of neutron star or black hole binaries with a population strongly peaking towards the Galactic Plane/Bulge (e.g. \citealt{nel04}). These studies suggest that about an equal fraction of semi-detached and detached systems are expected but the models over predict the number AM CVns observed in surveys like SDSS by at least a factor 10 \citep{roe07b,car13}, so the detached systems may well dominate. Most of the detached systems are predicted to consist of a carbon/oxygen + helium white dwarf binary system \citep{nel01b,nel13,rui10,liu10,yu10}. 

Although the currently known sample is still limited, upcoming and ongoing large scale high-cadence variability surveys which also cover low Galactic latitudes such as OmegaWhite (\citealt{mac15}), ZTF (\citealt{bel14}), BlackGEM (\citealt{bloem15}), GOTO (\citealt{ste17}), {\sl Gaia} and LSST (see \citealt{kor17} for both) have the potential to discover an unbiased sample of \emph{LISA} verification binaries. Indeed \citet{kor17} show that Gaia, LSST and \emph{LISA} have the potential to detect hundreds up to a few thousand new ultracompact double white dwarfs.

\section{Methods}\label{sec:method}
\subsection{Mass assumption for systems without constraints}
Mass estimation for AM\,CVn type systems are difficult because only the accretion disc and in some cases the accretor is visible in the spectra. So far the only AM\,CVn systems with direct measurements of the donor and the accretor mass are eclipsing systems. \citet{cop10} found precise masses for SDSS\,J092638.71+362402.4 and more recently \citet{gre18a} derived precise masses for the first fully eclipsing AM\,CVn type system: Gaia\,14aae. Both systems show a high accretor mass of 0.85 and 0.87\,\Msun\, respectively. Additionally, both systems have donor stars which are inconsistent with a zero-temperature fully degenerate star. In both cases the donor is larger and more massive compared to what it is expected for a fully degenerate donor. 

A large number of AM\,CVn systems have indirect constraints on the mass ratio ($q$) from the empirical relation of the superhump excess \citep{kni06}.
\begin{equation}
q = (0.114\pm0.005)+(3.97\pm0.41)\times(\epsilon-0.025) 
\end{equation}
where $\epsilon = \frac{P_{\rm sh}-P_{\rm orb}}{P_{\rm orb}}$ is the superhump excess. This is an empirical relation which gives similar results to that of \citet{pat05} but with the inclusion of uncertainties on the fit parameters. The relation was derived for hydrogen-dominated cataclysmic variables but has not yet been well tested for AM\,CVn type systems. \citet{gre18} applied the equation to 11 AM\,CVn systems with a measured superhump excess to derive the mass and radius for the donor under the assumption of an 0.7$\pm$0.1\,\Msun\, accretor. None of the tested systems are consistent with a fully degenerate donor but they are on average about 2.5 times the mass compared to a zero-temperature fully degenerate donor.

Based on this result and the measurements from the eclipsing systems we assume for systems without constraints on the component masses an accretor mass of $0.8\pm0.1$\,\Msun\, and a donor mass 2.5 times the mass for a zero-temperature donor star. For the donor star we allow an error range of $1.5-3.5$ times the minimum mass. System properties for each system are given in Table\,\ref{tab:system}.

\begin{table*}
\begin{center}
\caption{Measured EM properties (parallax, distance) and derived GW parameters (f, $\mathcal{A}$, signal-to-noise ratio SNR) of the known verification binaries. The distance for HM Cnc is assumed. The strain amplitude ($\mathcal{A}$) is given in units of $10^{-23}$. The SNR is calculated for four years integration with \emph{LISA}.} 
\begin{tabular}{lrrrrrrr}
\hline
Source                  &  $f$  &  $\varpi$ & $\sigma_\varpi$ & $d$ & $\sigma_d$ & $\mathcal{A}$  & SNR  \\
                        &  (mHz)   &  (mas)         & (mas)  & (pc) & (pc) &     &  \\
\hline
 \multicolumn{3}{l}{{\bf AM\,CVn type systems}}    &  &  &     \\
HM Cnc                      & 6.22  &  - & -           & [5000] & - & 6.4 & 211.1$\pm$3.18 \\
V407 Vul                    & 3.51  &  0.095  & 0.327  & 1786 & 667 & 11.0$\pm$5.9 & 169.7$\pm$2.17 \\
ES Cet                      & 3.22  &  0.596  & 0.108  & 1584 & 291 & 10.7$\pm$4.6 & 154.3$\pm$2.09 \\ 
SDSS\,J135154.46--064309.0  & 2.12  &  0.596  & 0.313  & 1317 & 531 &  6.2$\pm$3.5 & 21.8$\pm$0.24 \\
AM\,CVn                     & 1.94  &  3.351  & 0.045  &  299 &   4 & 28.3$\pm$3.2 & 101.2$\pm$0.96 \\
SDSS J190817.07+394036.4    & 1.84  &  0.954  & 0.046  & 1044 &  51 &  6.1$\pm$2.4 & 20.3$\pm$0.13 \\
HP Lib                      & 1.81  &  3.622  & 0.052  &  276 &   4 & 17.5$\pm$3.9 & 43.7$\pm$0.28 \\
PTF1 J191905.19+481506.2    & 1.48  &  0.550  & 0.327  & 1338 & 555 &  3.2$\pm$1.8 & 4.0$\pm$0.02 \\
CXOGBS J175107.6--294037    & 1.45  &  1.016  & 0.146  &  971 & 156 &  4.2$\pm$1.8 & 4.5$\pm$0.02  \\
CR Boo                      & 1.36  &  -  &  -  &  337$^a$  & ${^{+44}_{-35}}^a$   & 13.4$\pm$4.2  & 21.9$\pm$0.13\\
V803 Cen                    & 1.25  &  -  &  -  & 347$^a$ &  ${^{+32}_{-27}}^a$  & 16.0$\pm$5.4 &  26.2$\pm$0.17 \\
     \noalign{\smallskip}
 \multicolumn{3}{l}{{\bf detached white dwarfs}}    &   &   \\
SDSS J065133.34+284423.4    & 2.61  &  1.000 & 0.476   & 933  & 493 & 16.2$\pm$8.6 & 90.1$\pm$1.13\\
SDSS J093506.92+441107.0    & 1.68  &  - & -           & 645$^b$ & 41$^b$ & 29.9$\pm$7.7 & 44.9$\pm$0.31 \\
SDSS J163030.58+423305.7    & 0.84  & 0.937 & 0.270    & 1019 & 357 & 11.5$\pm$4.9 & 4.6$\pm$0.03 \\
SDSS J092345.59+302805.0    & 0.51  & 3.340 & 0.173    & 299  & 10  & 26.4$\pm$6.5 & 5.6$\pm$0.06 \\
      \noalign{\smallskip}
  \multicolumn{3}{l}{{\bf hot subdwarf binaries}}     &   &  \\
CD--30$^\circ$11223          & 0.47  & 2.963 & 0.080    & 337  &  9  & 41.5$\pm$1.8 & 4.9$\pm$0.04\\
\hline
\end{tabular}
\begin{flushleft}
$^a$\citet{roe07c}, $^b$\citet{bro16b}
\label{tab:system1}
\end{flushleft}
\end{center}
\end{table*}

\subsection{Distance determination from Gaia DR2 parallaxes}\label{sec:distance}
{\sl Gaia} DR2 provides parallaxes, not distances. In this Section we explain the procedure we adopt to convert parallaxes into distances.
To estimate distances from the measured parallaxes a probability-based inference approach is required \citep[e.g.][]{bai15,igo16,ast16,bai18,luri18}. Essentially, because any measured parallax ($\varpi$) follows a probability distribution, we can infer the distance in a probabilistic sense, if we make an assumption about the true distribution of observed sources in space (i.e. the prior distribution). Using Bayes' theorem the posterior probability density of the possible values for the distance can be expressed as

\begin{equation}
\begin{aligned}
P(d|\varpi,\sigma_{\varpi})= \frac{1}{Z}\  P(\varpi|d,\sigma_{\varpi})\  P(d);\\
Z = \int_{0}^{\infty} P(\varpi|r,\sigma_{\varpi}) \ P(r)\,d r,
\end{aligned}
\end{equation}
where $Z$ is the normalization constant, $P(\varpi|d,\sigma_{\varpi})$ is the likelihood function and $P(d)$ is the prior. The likelihood expresses the probability to measure the parallax $\varpi$ for the source at the distance $d$ with an uncertainty of the measurement $\sigma_{\varpi}$. For {\sl Gaia} measurements we can assume a Gaussian noise model \citep{lin18} and write the likelihood as
\begin{equation} \label{eqn:posterior}
P(\varpi|\ d,\sigma_{\varpi}) = \frac{1}{\sqrt{2\pi}\ \sigma_{\varpi}} \exp \left[{-\frac{1}{2\sigma_{\varpi}^2} \left( \varpi - \frac{1}{d} \right)^2} \right].
\end{equation}
The prior $P(d)$ contains our assumption about the distance distribution of the sources. For measurements with fractional parallax errors $\sigma_{\varpi}/\varpi$ less than about $0.1-0.2$, the distance estimates
are mainly independent of the choice of prior. However, for larger fractional errors the quality of the distance estimates heavily depends on how well the prior reflects the true distribution of distances for the population of sources \citep[e.g.][]{bai15, ast16}.
For this work we adopt an exponentially decreasing volume density prior

\begin{figure}
  \centering
  \includegraphics[width=0.48\textwidth]{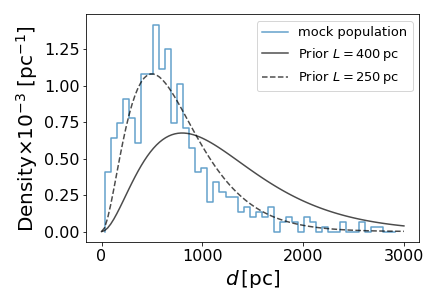}
 \caption{Distribution of synthetic detached double white dwarf binaries with  distance from \citet{kor17} (blue line) and two exponentially decreasing volume density priors: $L = 250\,$ pc (dashed black line) and $L=400\,$pc (solid black line). 
\label{fig:prior}
}
\end{figure}

\begin{figure*}
  \centering
  \includegraphics[width=0.9\textwidth]{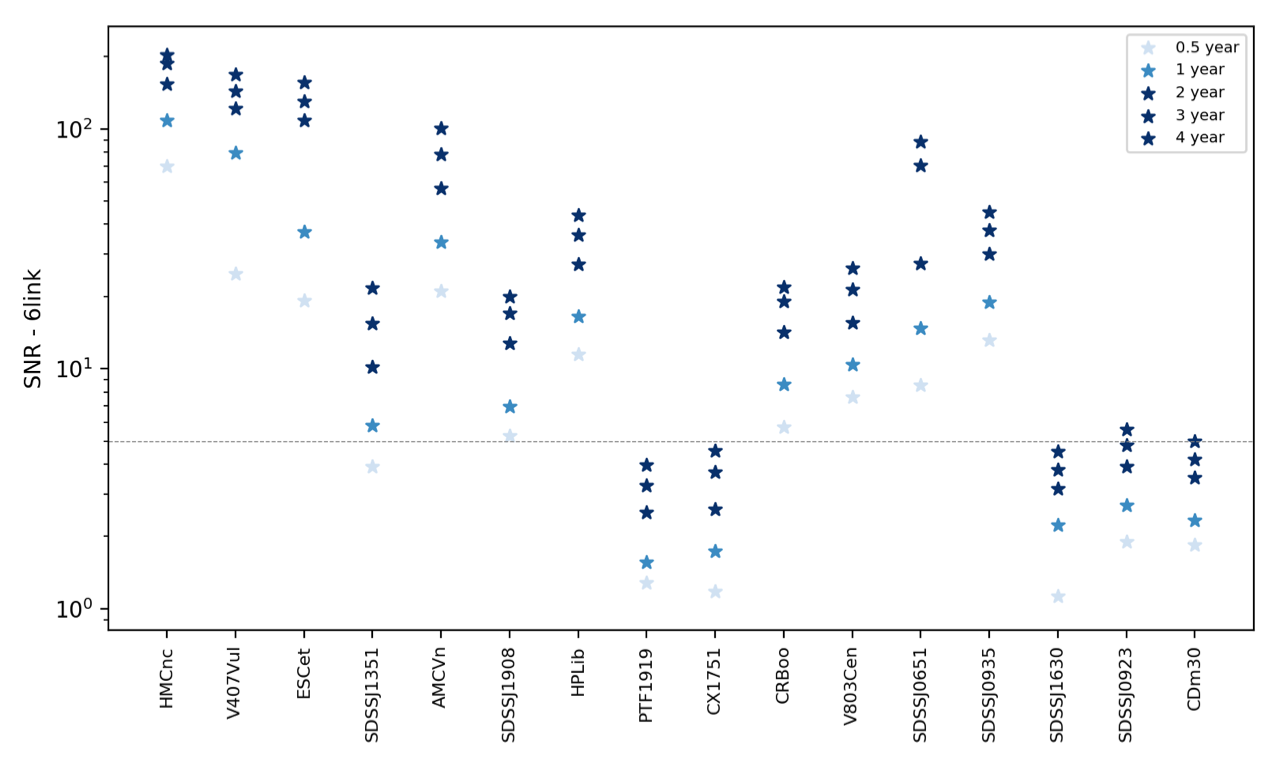}
 \caption{SNR evolution with time for the \emph{LISA} verification binaries. The black dashed line corresponds to SNR$=5$. 
\label{fig:SNRevol}
}
\end{figure*}

\begin{figure}
  \centering
  \includegraphics[width=0.48\textwidth]{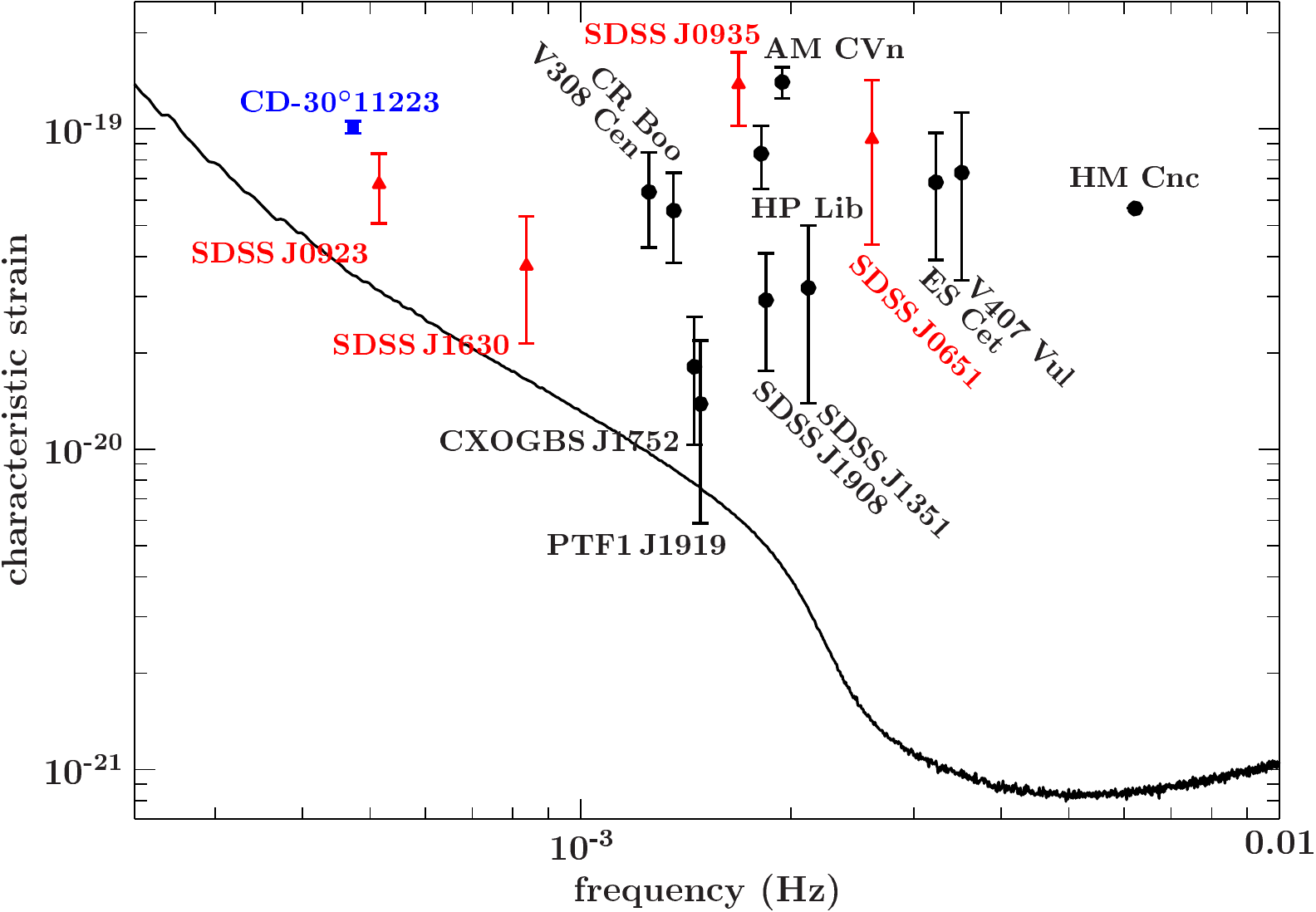}
 \caption{Sensitivity plot for \emph{LISA} adopted assuming 4 years integration from \citet{ama17} showing the verification binaries which reach a SNR$\geq5$ or are on the border to a SNR$\approx$5 after 4 years integration. Black circles are AM\,CVn systems, red triangles correspond to detached white dwarfs and the blue square is the hot subdwarf binary. Note that the gravitational frequency shown here is twice the orbital frequency of the binaries. We assume a distance of 5 kpc for HM Cnc. 
\label{fig:SNdata}
}
\end{figure}

\begin{equation}  \label{eqn:prior}
P(d) = \begin{cases}\frac{ d^2}{2L^3} \exp(-d/L) & \text {if} \  d > 0,\\
0 & \text{otherwise}, \end{cases}
\end{equation}
where $L > 0$ is the scale length. This prior performs well for a generic population, but fine-tuning is required to find the appropriate scale length L that describes \emph{LISA} verification binaries \citep{ast16}. We calibrate the value of $L$ using the mock catalog of detached double white dwarf binaries from \citet{kor17}. The catalog was obtained using the binary population synthesis model of \citet{nel01a,too17} and designed to test the detectability of these binaries by {\sl Gaia}, LSST and \emph{LISA}. We select binaries in the catalog with Gaia $G$ magnitude $<21$ and where parallax fractional error\footnote{The errors on parallax for the mock population are estimated using {\sc pyGaia} python tool kit.} $>0.2$. 
A straightforward way of fine-tuning the value of L is to fit the distribution of synthetic binaries with the distance. 
Another way consists of finding the value of $L$ that minimizes the bias on our estimates due to a particular choice for $L$ itself \citep[e.g.][]{mar18}. The latter implies the following calculations. For each binary we determine the mode of the posterior distribution $P(d|\varpi,\sigma_{\varpi})$. The mode is an unbiased estimator and provides meaningful estimates when the posterior is highly asymmetric. To determine the mode of $P(d|\varpi,\sigma_{\varpi})$  given our choice of the prior, we set the derivative of the posterior to be equal to zero and solve the equation \citep{bai15}:  
\begin{equation} \label{eqn:mode}
\frac{d^3}{L}-2d^2 + \frac{\varpi d}{\sigma_{\varpi}^2}-\frac{1}{\sigma_{\varpi}^2} = 0.
\end{equation}
We repeat this calculation using a range of values for $L$ for each binary in the mock catalog.
We define the best value of $L$ as the one that minimizes the difference between the mode, obtained by solving Eq.~\eqref{eqn:mode}, and the true distance of the binary in the catalog. We obtain $L=400\,$pc. 
In Figure\,\ref{fig:prior} we show the distribution of synthetic binaries with the distance (blue line) and two exponentially decreasing  volume density priors: one with $L = 250\,$ pc  (dashed black line), that represents the best fit to the distribution of mock binaries, and another one with $L=400\,$pc (solid black line), obtained by minimazing the bias. The figure shows that the curve with $L=400\,$pc decreases slower and is more representative of binaries at large distances, where fractional errors on parallax are large. Thus, for this work we adopt the scale length of $400\,$pc such that we avoid underestimating distances for the furthest binaries.
Finally, following \citet{bai15} we associate the most probable value of $d$ with the mode of the posterior distribution, and we compute the errors as
\begin{equation}
\sigma_d = \frac{d_{95}-d_5}{2s},
\end{equation}
where $d_{95}$ and $d_5$ are the boundaries of the 90\% credible interval of the $P(d|\varpi,\sigma_{\varpi})$ distribution that are calculated symmetrically about the median and $s=1.645$, which is the ratio of the 90\% to 68.3\% credible interval for a Gaussian distribution. HM\,Cnc, CR\,Boo, V803\,Cen and SDSS J093506.92+441107.0 have no measured parallax from Gaia DR2. For HM\,Cnc we assumed 5 kpc and discuss the uncertainty on the distance in detail in \S \ref{sec:discussion}. For the other three systems the previously published distant estimates were used. The results are listed in Table\,\ref{tab:system1}. 

\begin{figure*}
  \centering
  \includegraphics[width=0.9\textwidth]{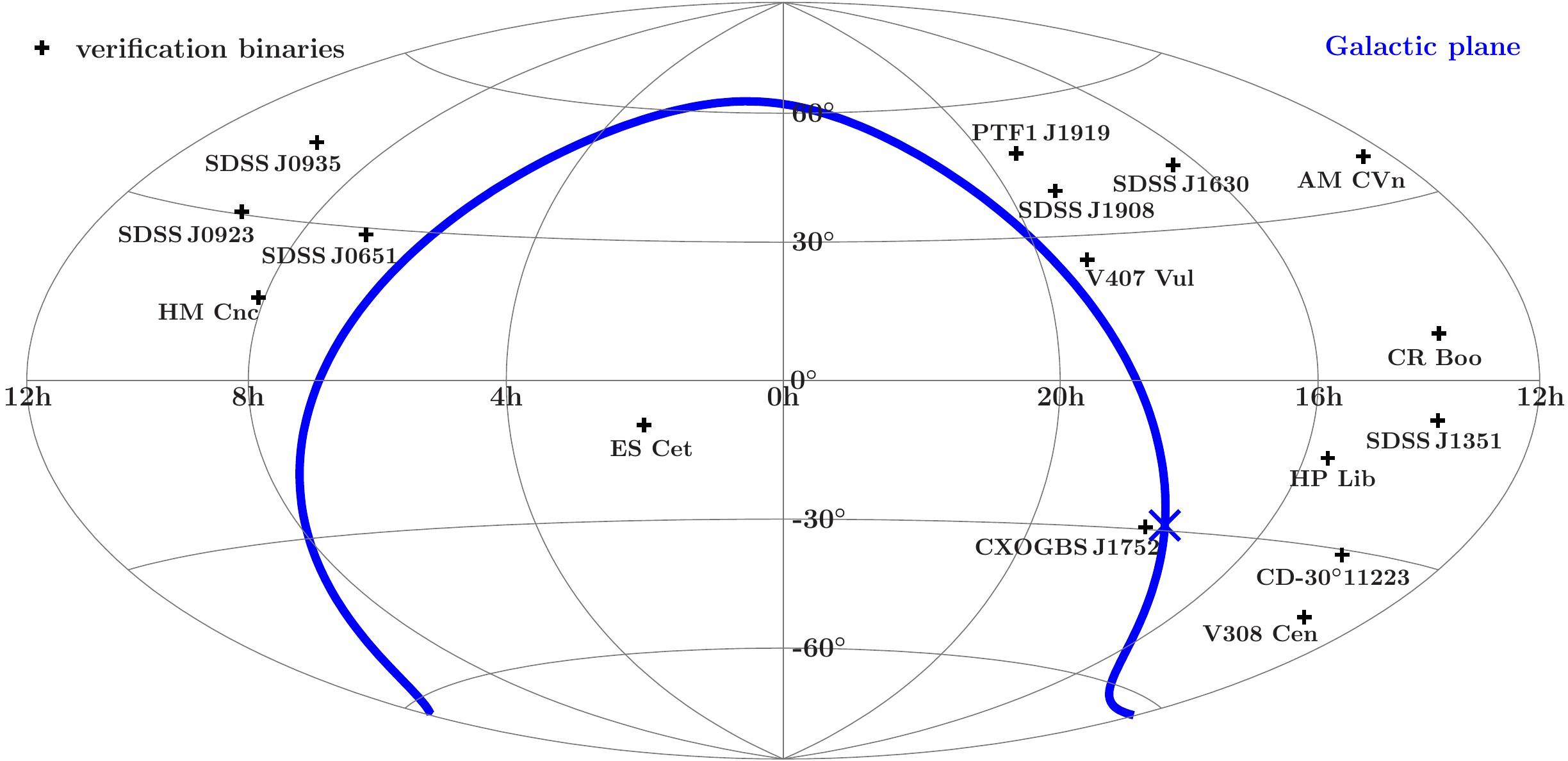}
 \caption{Sky position of the verification binaries. The sky positions show a clear bias towards the Northern hemisphere and to higher Galactic latitudes. The blue line indicates the Galactic Plane, with the Galactic Center located at the blue cross.
 }
  \label{fig:map}
\end{figure*}

\subsection{Strain and SNR calculations}\label{sec:snrcal}
To compute the expected characteristic strain we first calculate the dimensionless gravitational wave amplitude ($\mathcal{A}$) using equation 3 from \citet{sha12}:
\begin{equation}
\mathcal{A} = \frac{2(G\mathcal{M})^{5/3}}{c^4d}(\pi f)^{2/3}
\end{equation}
where $\mathcal{M}$ is the chirp mass, $\mathcal{M} \equiv (m_1m_2)^{3/5}/(m_1+m_2)^{1/5}$, $m_1$ and $m_2$ the masses of the two components, where we assume $m_1>m_2$, $d$ is the distance to the source as defined in \S \ref{sec:distance} and $f$ the gravitational wave frequency with $f=2/P_{\mathrm{orb}}$.
The characteristic strain ($h_c$) for individual verification binaries was calculated following the approach described in Section 2 in \citet{moo14}: 
\begin{equation}
h_c = \sqrt{N_{\rm cycle}}\mathcal{A}
\end{equation}
where $N_{\rm cycle}=fT_{\rm obs}$. For the calculations we assume that \emph{LISA} will observe for four years. The masses and gravitational wave frequency for each system are given in Tables\,\ref{tab:system} and \ref{tab:system1}.

Most of the \emph{LISA} verification binaries can be characterized as monochromatic GW signals with a set of seven parameters, $\mathcal{A}$, $f$, polarization angle ($\psi$), initial GW phase ($\phi_{0}$), orbital inclination ($\iota$), ecliptic latitude ($\sin\beta$), and ecliptic longitude ($\lambda$). An additional eighth parameter, the period derivative or chirp ($\dot{f}$) is used for HM Cnc, V407~Vul and SDSS\,J0651 which have measured orbital decay rates from their EM data, $\dot{P}_{\mathrm{orb}}$:\ $3.75\times 10^{-11} s s^{-1}$ (\citealt{roe10}), $3.17\times 10^{-12} s s^{-1}$ (\citealt{ram05}), $9.8 \pm 2.8 \times 10^{-12} s s^{-1}$ (\citealt{her12}) respectively. They are related to the GW decay rate, $\dot{f}$ by $-\dot{P}_{\mathrm{orb}}/P_{\mathrm{orb}}^2$ used in simulating their GW signals. We compute Fisher matrices (e.g \citealt{cut98}) to extract the GW parameter uncertainties and correlations. The method and application of Fisher information matrix (FIM) for the \emph{LISA} compact binaries together with their signal modeling and the noise from the detector and the Galactic foreground have been described in detail in \cite{sha12}. 

Here we use the current configuration for the \emph{LISA} detector (\citealt{ama17}) with armlengths of $2.5 \times 10^{6}$ km and six laser links exchanged along the three arms of the detector, from which we can generate two sets of the optimal data streams from two channels yielding two independent time-series whose noises are uncorrelated maximizing the SNR\footnote{We use the Time Delay Interferometry (TDI) A and E observables}. Detailed discussions of the possible data streams using various sets of laser links can be found in e.g \cite{val05}. The unresolved foreground is obtained by using the recently updated catalog for detached double white dwarf binaries whose simulation and binary evolution is described in \citet{too17}. 

We obtain the SNR from the GW signal over 15 instrumental noise realizations for the bright verification binaries using the nominal EM measurements to estimate the GW parameters in the GW signal model. For details we refer to \S 3 in \cite{sha12}. Given the GW signal of the binary and a Gaussian noise we can use FIM to estimate the parameter uncertainties. The inverse of the FIM is the variance-covariance matrix whose diagonal elements are the GW uncertainties and the off-diagonal elements are the correlations between the two parameters. We do the GW analysis of the above mentioned verification binaries for \emph{LISA} observations of four years. We note that the Fisher-based method is a quick way of computing  parameter uncertainties and their correlations in which these uncertainties are estimated locally at the true parameter values and therefore by definition the method cannot be used to sample the entire posterior distribution of the parameters. Additionally Fisher-based results hold in the limit of strong signals with a Gaussian noise (see the Appendix in \citealt{sha14})\footnote{The code used to perform the simulation is available at https://doi.org/10.17617/1.68}.

\begin{table*}
\centering
\caption{GW parameter uncertainties for the bright verification binaries from Fisher Information Matrix after four years of LISA integration}
\label{tab:VBs_fisher_errors}
\begin{tabular}{lrrrr}
\hline \hline
Source &SNR&$\sigma_{\mathcal{A}/\mathcal{A}}$ & $\sigma_{\iota}[^\circ]$ & $c_{\mathcal{A} \:\iota}$ \\
\hline
HM Cnc                      & 211.1$\pm$3.18 & 0.07$\pm$0.001  & 5.82$\pm$0.08  & 0.991$\pm$0.029\\
V407 Vul                    & 169.7$\pm$2.17 & 0.028$\pm$0.000  & 1.34$\pm$0.02 & 0.907$\pm$0.023 \\
ES Cet                      & 154.3$\pm$2.09 & 0.032$\pm$0.000  & 1.44$\pm$0.02 & 0.911$\pm$0.024\\
SDSS\,J135154.46--064309.0  & 21.8$\pm$0.24 & 0.218$\pm$0.002  & 10.22$\pm$0.11 &  0.911$\pm$0.020\\
AM\,CVn                     & 101.2$\pm$0.96 & 0.113$\pm$0.001  & 8.03$\pm$0.08  & 0.985$\pm$0.018\\
SDSS J190817.07+394036.4    & 20.3$\pm$0.13 & 5.622$\pm$0.036 & --$^a$ &  1.000$\pm$0.013\\
HP Lib                      & 43.7$\pm$0.28 & 0.599$\pm$0.004  & 63.82$\pm$0.41 &  0.997$\pm$0.013\\
PTF1 J191905.19+481506.2    & 4.0$\pm$0.02  & 1.218$\pm$0.008  & 57.54$\pm$0.33 &  0.909$\pm$0.011\\
CXOGBS J175107.6--294037    & 4.5$\pm$0.02  & 1.057$\pm$0.005 & 49.65$\pm$0.33 &  0.909$\pm$0.009\\
CR Boo                      & 21.9$\pm$0.13 & 1.173$\pm$0.007  & 126.05$\pm$0.72 & 0.997$\pm$0.011\\
V803 Cen                    & 26.2$\pm$0.17 & 4.647$\pm$0.029 & --$^a$ &  1.000$\pm$0.013\\
SDSS J065133.34+284423.4    & 90.1$\pm$1.13 & 0.022$\pm$0.000  & 0.65$\pm$0.01 & 0.159$\pm$0.004\\
SDSS J092345.59+302805.0    & 44.9$\pm$0.31 & 0.106$\pm$0.001  & 4.99$\pm$0.03 & 0.909$\pm$0.013\\
SDSS J163030.58+423305.7    & 4.6$\pm$0.03 & 1.064$\pm$0.008  & 49.29$\pm$0.39 &  0.909$\pm$0.014\\
SDSS J092345.59+302805.0    & 5.6$\pm$0.06 & 0.834$\pm$0.009  & 39.51$\pm$0.44 & 0.908$\pm$0.020\\
CD--30$^\circ$11223          & 4.9$\pm$0.04 & 0.425$\pm$0.004  & 12.52$\pm$0.13 &  0.359$\pm$0.007\\
\hline
\end{tabular}
\begin{flushleft}
$^a$The FIM uncertainty exceeds the physically allowed range by $\iota$ and thus cannot be determined from GW data analysis \citet{sha14}
\end{flushleft}
\end{table*}

\section{Results}

We calculate the distance and expected SNR following the description outlined in \S \ref{sec:method} for $\approx$50 semi-detached and detached candidate verification binaries with the strongest expected gravitational wave signals. Table\,\ref{tab:system1} presents the predicted gravitational wave amplitude ($\mathcal{A}$) as well as the expected SNR after 4 years integration with \emph{LISA} for all systems with SNR$\geq$5 and systems which are on the border to a SNR$\approx5$.
 
We find that 13 systems reach a SNR$\geq$5 after four years observing with \emph{LISA} and therefore are confirmed verification binaries based on the definition adopted in \S \ref{sec:intro}. The population consists of 9 AM\,CVn binaries: HM\,Cnc, V407\,Vul, ES\,Cet, SDSS\,J1351, AM\,CVn, SDSS\,J1908, HP\,Lib, CR\,Boo and V803\,Cen, 3 double white dwarfs: SDSS\,J0651, SDSS\,J0935 and SDSS\,J0923 as well as CD--30$^\circ$11223 the first verification binary consisting of a hot subdwarf star with a massive white dwarf companion. Additionally, we find three more systems (PTF1\,J1919, CXOGBS\,J1751 and SDSS\,J1630) with a SNR of $\approx$5 or just below $5$, making them good candidates for being verification binaries.

Figure\,\ref{fig:SNRevol} shows the evolution of the SNR after 0.5, 1, 2, 3 and 4 years. The loudest source is HM\,Cnc which reaches a SNR=211 after four years of integration and already SNR=69 after 0.5 years, but whose distance is, even after \emph{Gaia} DR2, still poorly constrained. PTF1\,J1919, CXOGBS\,J1751, SDSS\,J1630, SDSS\,J0923 and CD--30$^\circ$11223 need four years of integration to reach a SNR$\approx$5. Figure\,\ref{fig:SNdata} shows the \emph{LISA} sensitivity curve and the characteristic strain of the verification binaries after four years observing with \emph{LISA}. 

Table\,\ref{tab:VBs_fisher_errors} shows the parameter uncertainties extracted from the FIM. Of the seven GW parameters characterizing a binary, the astrophysically interesting ones are the amplitude ($\mathcal{A}$) and the inclination ($\iota$). Shown are the relative $1-\sigma$ error in $\mathcal{A}$, absolute $1-\sigma$ error in $\iota$ and the normalized correlation between the two parameters $c_{\mathcal{A} \:\iota}$. The SNR influences the parameter uncertainties. Then the correlation $c_{\mathcal{A} \:\iota}$ has a strong influence on their uncertainties (\citealt{sha12}). As a result the systems with lower inclinations (or face-on orientations) with $\iota = [0^\circ - 45^\circ]$ have a strong correlation due to the GW signals being indistinguishable by making changes in $\mathcal{A}$ or $\iota$. This explains the large errors in $\mathcal{A}$ and undetermined $\iota$ (since the GW uncertainties are greater than the physical values $\iota$ can take: $0^\circ-360^\circ$) for systems such as SDSS J190817.07+394036.4, CR Boo and V803 Cen despite that their SNRs are greater than 20. Whereas CD--30$^\circ$11223 has a better constrained inclination even though its SNR is lower at $\approx$5.

\section{Discussion}\label{sec:discussion}

The only remaining system without a distance measurement is HM\,Cnc: distance is therefore the largest uncertainty when predicting its $\mathcal{A}$ and SNR for \emph{LISA}. Given its known properties we argue that 10\,kpc is the most conservative estimation for the distance. Although $\mathcal{A}$ and SNR remain uncertain, we find that even at a distance of 10\,kpc, HM\,Cnc will have a SNR$\approx$100 after four years observing with \emph{LISA}. Hence it remains a bright verification binary even if the distance is significantly greater than the assumed 5\,kpc.

Although there is a {\sl Gaia} parallax measurement for V407 Vul (0.095$\pm$0.327), the optical counterpart is dominated by a component that matches a G-type star, with a blue variable component in phase with the binary making up only 10-40\% of the flux (\citealt{ste06}). The probability that this is an unrelated chance alignment of a foreground object is small, but the specific association of this G-star component with the ultra-compact binary is unclear. Given that this star dominates the Gaia passband, we assume here that the parallax measurement for V407 Vul is essentially that of the G-star component, and adopt this also for the ultra-compact binary component given the close on-sky alignment. 


With the current \emph{LISA} configuration and four years of observation, $> 27,000$\footnote{Estimate from the foreground simulation using \citet{too17} catalogue} binaries are expected to be individually detected by \emph{LISA}. However, only a small fraction will be bright enough to be detectable in the optical. \citet{nel04} predict that 143 short period semi-detached \emph{LISA} verification systems (3 in the direct-impact phase) with P$_{\rm orb}<1500$\,s and brighter than 20\,mag should be detectable in the optical wavebands. More recent work by \citet{kor17} predict several tens, up to one hundred, detached double white dwarfs will be detectable in the optical bands by {\sl Gaia} and LSST as eclipsing sources, those with high SNR from their gravitational waves and brighter than 24 mag. The eclipsing systems only represent a small fraction of the full sample and we expect that there are about 100 detached double white dwarfs with orbital periods below 10\,min and brighter than 24 mag and therefore potentially detectable with LSST. 

Since verification binaries are a Galactic population their surface density is expected to strongly peak near the Galactic Plane. Most of the known systems are located in the Northern hemisphere and only a few systems were found at low Galactic latitudes. This shows that the current sample is likely very incomplete and biased. Figure\,\ref{fig:map} shows the sky position of the 16 systems. Upcoming and ongoing large scale optical surveys such as OmegaWhite (\citealt{mac15}), ZTF (\citealt{bel14}), BlackGEM (\citealt{bloem15}), GOTO (\citealt{ste17}), {\sl Gaia} and LSST (see \citealt{kor17} for both) are expected to discover a more unbiased sample across both hemispheres and at low Galactic latitudes before \emph{LISA} gets launched.

\section{Summary and Conclusions}

In this work we derived distances from {\sl Gaia} DR2 parallaxes for $\approx$50 verification binary candidates. Using these distances, we  calculated the expected SNR after four years integration with \emph{LISA} with a configuration of 6 laser links and 2.5 Gm arm lengths. Given the definition of a verification binary as SNR$\geq$5 after four years integration, we find a total of 13 verification binaries. Eleven systems reach a SNR$\geq$20 and two additional systems reach a SNR$\geq$5 after four years. Additionally we find three more systems which are  expected to have a SNR$\approx$5 after four years integration with \emph{LISA} and are good candidates for being verification binaries. Our study confirmed the first hot subdwarf binary as a \emph{LISA} verification binary.

So far, distances have been the most uncertain parameter when predicting the gravitational wave strengths of the bright verification binaries. This is in particular true for the systems with the most accurate constraints on system parameters such as masses, inclinations and orbital periods. We find that {\sl Gaia} provides accurate distances in particular for systems which are at most a few hundred parsec away. This allows us to predict the gravitational wave amplitude ($\mathcal{A}$) with an accuracy better than 5\,\% in the case of CD--30$^\circ$11223 and around 10\,\% for AM\,CVn itself, making these systems ideal for the performance validation of \emph{LISA}. For the remaining systems with distances of a few hundred parsec (e.g. HP\,Lib and SDSS\,J0923), the uncertainty of the gravitational wave amplitude is now dominated by the uncertainty on the component masses. For these systems and future discoveries precise mass measurements are required to provide estimations on the gravitational wave strength with a precision of a few percent.




\section*{Acknowledgments}
TK would like to thank Thomas Tauris for useful comments on the manuscript. VK would like to thank Tommaso Marchetti for useful discussion on derivation of Gaia distances. This work presents results from the European Space Agency (ESA) space mission Gaia. Gaia data is being processed by the Gaia Data Processing and Analysis Consortium (DPAC). Funding for the DPAC is provided by national institutions, in particular the institutions participating in the Gaia MultiLateral Agreement (MLA). The Gaia mission website is \url{https://www.cosmos.esa.int/gaia}. The Gaia archive website is \url{https://archives.esac.esa.int/gaia}.
Armagh Observatory and Planetarium is core funded by the Northern Ireland Executive through the Dept. for Communities. This research made use of NumPy \citep{van2011numpy} This research made use of matplotlib, a Python library for publication quality graphics \citep{Hunter:2007} This research made use of Astropy, a community-developed core Python package for Astronomy \citep{2013A&A...558A..33A} 

\bibliographystyle{mnras}
\bibliography{refs,refs_1508}

\label{lastpage}
\end{document}